\documentclass[a4paper,english]{article}
\usepackage{pslatex}
\usepackage[T1]{fontenc}
\usepackage[latin1]{inputenc}
\usepackage{graphicx}

\makeatletter

\providecommand{\LyX}{L\kern-.1667em\lower.25em\hbox{Y}\kern-.125emX\@}
\usepackage{graphicx}

\def\apj{{\it Astrophys. J.\ }}
\def\apjl{{\it Astrophys. J. Lett.\ }}

\def\aap{{\it Astron. Astrophys.\ }}

\def\mnras{{\it Mon. Not. R. Astron. Soc.\ }}

\def\proc#1{{\it Proc.\ #1th Int. Cosmic Ray Conf.\ }}

\def\rpp{{Rep.~Progr.~Phys.\ }}
\def\ass{{Astroph Space Sci\ }}
\def\mathcal{{\it }}

\def\etal{{\it et al.,\ }} % macros exists
\def\eg{{ e.g.,\ }}

\def\ie{{ i.e.,\ }}

\def\ln{{\rm ln}}

\usepackage{hyperref}

\makeatother

\begin{document}
\onecolumn

\setlength{\baselineskip}{17.5pt}

\vspace*{-0.65in}

\begin{center}\textbf{\large Generation of Mesoscale Magnetic Fields
and the Dynamics of Cosmic Ray Acceleration }\end{center}{\large \par}

\begin{center}\vspace*{0.30in}\end{center}

\begin{center} DIAMOND, Patrick H. and  MALKOV, Mikhail A.\end{center}

\begin{center}\textit{University of California at San Diego, La Jolla,
California 92093-0319, USA }\end{center}

\begin{center}\textit{e-mail: pdiamond@ucsd.edu}\end{center}

\vspace*{0.00in}

\noindent \textbf{Abstract}

\hspace*{12pt}

The problem of the cosmic ray origin is discussed in connection with
their acceleration in supernova remnant shocks. The diffusive shock
acceleration mechanism is reviewed and its potential to accelerate
particles to the maximum energy of (presumably) galactic cosmic rays
($10^{18}eV)$ is considered. It is argued that to reach such energies,
a strong magnetic field at scales larger than the particle gyroradius
must be created as a result of the acceleration process, itself. One
specific mechanism suggested here is based on the generation of Alfven
wave at the gyroradius scale with a subsequent transfer to longer
scales via interaction with strong acoustic turbulence in the shock
precursor. The acoustic turbulence in turn, may be generated by Drury
instability or by parametric instability of the Alfven waves.

\vspace*{0.20in}


\section{Introduction}

\begin{figure}
\includegraphics[  bb=24bp 90bp 517bp 661bp,
  clip,
  width=12cm,
  height=10cm]{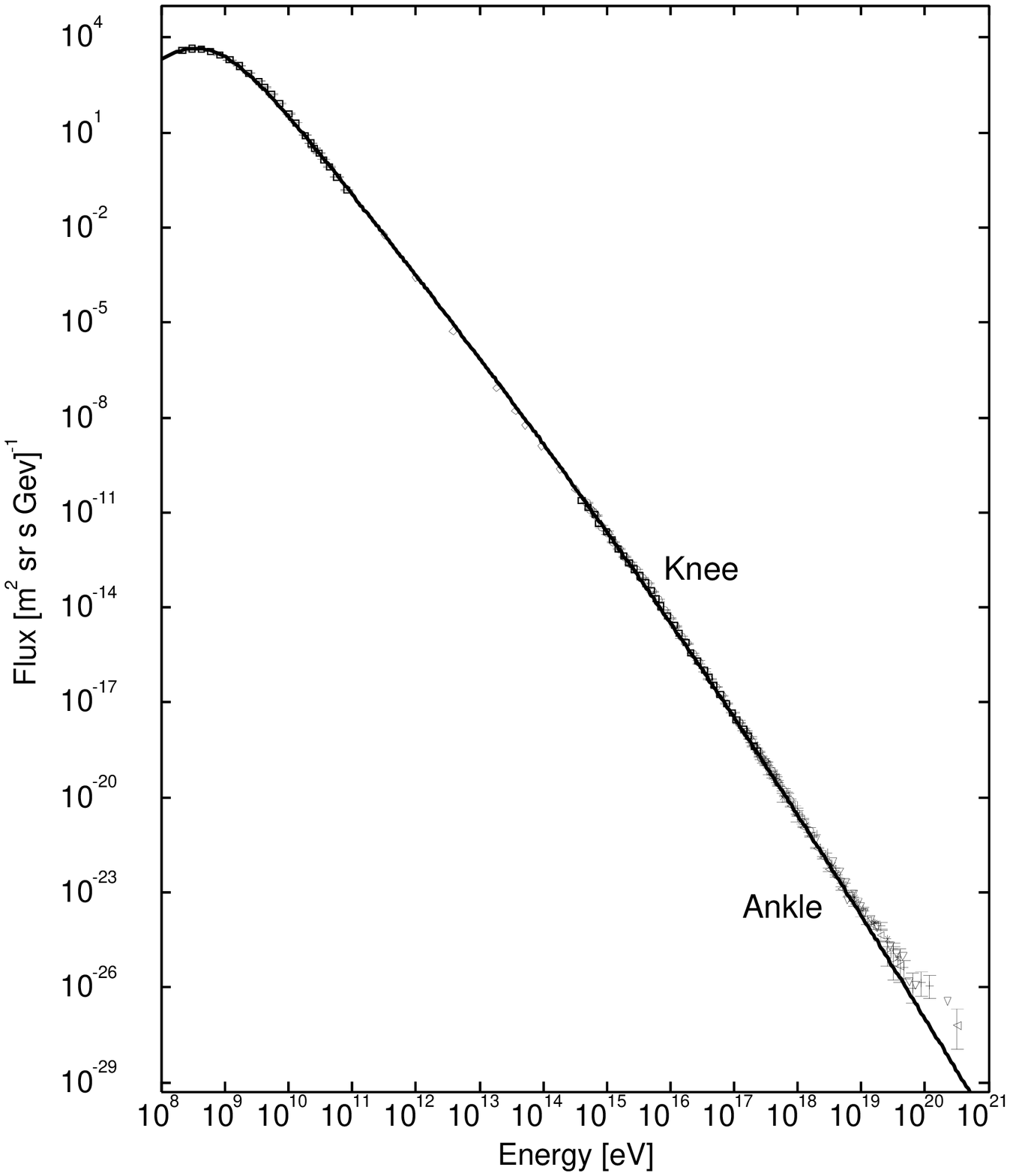}

\caption{Cosmic Ray spectrum \label{fig:1}}
\end{figure}

Cosmic rays (CRs) were discovered almost a century ago, yet their
origin is unknown. There is sufficient evidence that at least part
of their spectrum ($E<10^{18}eV$) originates in the Galaxy, while
particles of higher energies are thought to come to us from extragalactic
sources. One simple theoretical argument is that particles with such
high energies (usually referred to as ultra high energy CRs, UHECR)
could not be confined to the Galaxy, on account of their large gyroradius.
Second, their (power law- $E^{-2.7}$) spectrum is harder than that
of (presumably) galactic particles at $E<10^{18}eV$ ($E^{-3.1})$,
which is consistent with their extra-galactic origin. The last point
becomes clear if we turn to the other break on the overall CR spectrum,
the one at $E\sim 10^{15}eV$, commonly known as the {}``knee'',
shown in Fig.\ref{fig:1}. The spectrum above this energy steepens
(from $E^{-2.7}$ to $E^{-3.1}$), so that the premise of their extragalactic
origin would require an explanation of why the galactic part of the
spectrum terminates, while the extragalactic part appears \emph{exactly}
at the knee energy. As we shall see, to explain the cosmic ray power
law spectrum between the {}``knee'' at$10^{15}eV$ and the {}``ankle''
at $10^{18}eV$ in terms of acceleration within the Galaxy is one
of the most serious challenges of contemporary acceleration mechanisms
and one of the main motivations of this study.

The explanation of the spectrum beyond the ankle (the highest energy
event observed so far is $3\cdot 10^{20}eV$) poses a major challenge
to fundamental physics. Given the distance of possible accelerators
(at least a few tens of Mpc), particles of such high energy should
have experienced significant losses through their interaction with
the cosmic microwave background radiation (the so called Greisen,
Zatsepin, Kusmin or GZK cut-off) while propagating over long distances.
Note that UHECR are sub-atomic particles with the energy of a well-hit
baseball. This energy exceeds (by at least three orders of magnitude)
that achievable by all existing terrestrial accelerators (\eg  Large
Hadron Collider, LHC-$10^{17}eV$). Another fundamental aspect of
the problem of CR origin is that they are the ultimate receptacle
of a significant portion of the gravitational energy in the Universe.
Indeed, star formation from the gravitational collapse of the primordial
gas with subsequent SN (supernova) explosions and their blast waves
results in CR acceleration. On the observational side, there is an
evidence \cite{koyama,tanim98,allen01} (in the form of both synchrotron
and inverse Compton radiation) that electrons of energies up to $100TeV$
are accelerated in the supernova shock waves.

\section{Acceleration mechanism}

The leading CR acceleration mechanism, namely the diffusive shock
acceleration (DSA, also known as the 1st -order Fermi mechanism) was
proposed originally by Fermi in Ref.\cite{Fermi}, and in its modern
form by a number of authors in the late seventies \cite{kr77,axf77,bla:ost,bell78a}.
The mechanism is basically simple -- particles gain energy by bouncing
between converging upstream and downstream regions of the flow near
a shock wave such as that from an SNR (supernova remnant) shock. This
mechanism requires magnetic fields. First, the field binds particles
to the accelerator (shock wave), in the direction perpendicular to
the field. Confinement in the direction \emph{along} the field lines
is, in turn, ensured by particles themselves through the generation
of Alfven waves by accelerated particles streaming ahead of the shock.
This occurs via Doppler resonance $\Omega =k(p/m)\mu $, where $\Omega $
and $p$ are the (nonrelativistic) gyrofrequency and momentum, $k$
is the wave number, $m$ is the particle mass, and $\mu $ is the
cosine of its pitch angle. These waves, in turn, scatter particles
in pitch angle (at the rate $\nu \sim \Omega (mc/p)\left(\delta B/B_{0}\right)^{2}$,
where $c$ is the speed of light) back and force so that they can
gain energy by repeatedly crossing the shock. Particle self-confinement
along the field is thus diffusive and the diffusivity $\kappa \sim c^{2}/\nu $
is inversely proportional to the fluctuation energy $\delta \mathbf{B}^{2}$,
as the fluctuating field is responsible for pitch-angle scattering.
However, the mean field $B_{0}$ ultimately determines the acceleration
rate and the particle maximum energy since it sets the work done by
the induced electric field on the particles. Indeed, $E_{max}\sim (e/c)u_{s}B_{0}R_{s}$,
where $u_{s}$ and $R_{s}$ are the speed and typical size of the
shock wave, such as the radius of the SNR shock. The fluctuating part,
$\delta B$, is typically assumed to be saturated at most at the level
$\delta B\sim B_{0}$, which provides pitch-angle scattering at the
rate of gyrofrequency, and thus limits the particle mean free path
(m.f.p.) along the field to a distance of the order of gyroradius
(the so called Bohm diffusion limit). An important thing to keep in
mind is that, due to the resonance condition $kp=const$, confinement
of particles of higher energies requires that longer waves need be
excited.

The most critical test of this mechanism is the requirement that it
accelerates galactic CRs to the energy of $10^{15}$ eV over the life-time
of supernova remnant shocks. Even with the above {}``optimistic''
estimates of the turbulence level, the mechanism passes this test
at best only marginally. If the turbulence level is lower, then the
maximum energy should be reduced proportionally. There are indeed
a number of phenomena which may either reduce the turbulence level
\cite{lc83}, or which can shift the turbulence spectrum (in wave
number) away from resonance with the high energy particles and therefore
cause their losses \cite{mdj02}.

Another reason for concern about this mechanism, at least in its standard
(Bohm limit) version, is its seeming inability to explain acceleration
of particles beyond $10^{15}$eV. As was discussed above, the cosmic
ray spectrum has only a break at this energy, and continues to about
$10^{18}$ eV where the extragalactic component is believed to start
dominating the spectrum. 

One approach to this problem is to generate a fluctuating component
$\delta B$ significantly exceeding the unperturbed field $B_{0}$
\cite{lucbel}. Physically, such generation is possible since the
free energy source is the pressure gradient of accelerated particles,
which in turn may reach a significant fraction of the shock ram energy.
Specifically, the wave energy density $\left(\delta B/B_{0}\right)^{2}$
may be related to the partial pressure $P_{c}$ of CRs that resonantly
drives these waves through the relation \cite{vdm84}

\begin{equation}
\left(\delta B/B_{0}\right)^{2}\sim M_{A}P_{c}/\rho u_{s}^{2}\label{delB}\end{equation}
 where $M_{A}=u_{s}/V_{A}\gg 1$ is the Alfven Mach number and $\rho u_{s}^{2}$
is the shock ram pressure. Of course, when $\delta B/B_{0}$ exceeds
unity, particle dynamics, and thus their confinement and acceleration
rates, are very difficult to assess if the turbulence spectrum is
sufficiently broad. The numerical studies by \cite{lucbel} showed
that at least in the case of an MHD (magnetohydrodynamic) description
of the background plasma and rather narrow wave (and particle energy)
band, the amplitude of the principal mode can reach a few times that
of the background field. Moreover, the authors of Ref.\cite{belluc01}
argue that in the case of efficient acceleration, field amplification
may be even stronger, reaching a $mG$ ($10^{-3}$ Gauss) level from
the background of a few $\mu G$ ($10^{-6}$ Gauss) ISM field, thus
providing acceleration of protons up to $10^{17}$ eV in SNRs.

Recently, the authors of Ref.\cite{ptuskin03} approached this problem
from a different perspective. They considered a Kolmogorov turbulent
cascade to small scales assuming the waves are generated by efficiently
accelerated particles on the long-wave part of the spectrum. They
obtained a particle maximum energy similar to that of \cite{belluc01}. 

Apart from the excitation of magnetic fluctuations during acceleration
process, there is yet another aspect of the CR-- magnetic field connection
discussed in the literature. Zweibel\cite{zweibel03} points out that
since CRs were already present in young galaxies (observed at high
redshifts), magnetic fields of appreciable strength must also have
been there at that time. She emphasizes, however, that the approximate
equipartition between the CR and magnetic field energy established
by the current epoch in our Galaxy is not required for the acceleration
mechanism and presumably results from the fact that they both have
a common energy source, namely the supernovae. Indeed, as we discussed
already, the magnetic field strength merely determines the \emph{maximum}
energy of accelerated particles, given the time available for acceleration
and the size of the accelerator. If the latter are sufficient then
the total energy of accelerated particles can, and in most of the
DSA models \emph{does,} exceed that of the magnetic field. The latter,
in turn, remains unchanged, apart from the conventional compression
at the shock and the MHD fluctuations discussed earlier.

In this paper we discuss the possibility of a \emph{different} scenario,
in which the magnetic field may absorb a significant part of the shock
energy as a result of the acceleration process, which may in fact
be strongly enhanced. The mechanism of such enhancement is based on
the transfer of magnetic energy to longer scales, which we call \emph{inverse
cascade} for short, even though specific mechanisms of such transfer
may differ from what is usually understood as a cascade in MHD turbulence.
This transfer is limited only by some outer scale $L_{out}$ such
as the shock precursor size $\kappa (p_{max})/u_{s}\sim r_{g}(p_{max})c/u_{s}\gg r_{g}(p_{max})$.
This approach is in contrast to the above discussed models \cite{belluc01,ptuskin03},
which operate on the generated magnetic fields with the scale lengths
of the order of the Larmor radius $r_{g}(p_{max})$ of the highest
energy particles and smaller. The advantage of the inverse cascade
for the acceleration is that the turbulent field at the outer scale
$\delta B(L_{out})\equiv B_{rms}$ (which necessarily must have long
autocorrelation time) can be obviously regarded as an {}``ambient
field'' for accelerated particles of all energies. If $B_{rms}\gg B_{0}$,
then the acceleration can be enhanced by a factor $B_{rms}/B_{0}$.
Note that the resonance field $\delta B(r_{g})$ may remain smaller
than $B_{rms}$, so that standard arguments about Bohm diffusion apply,
and it is less likely that this field will be rapidly dissipated by
nonlinear processes, such as induced scattering on thermal protons,
not included in the enhanced acceleration model \cite{lucbel,belluc01}.

As it should be clear from the above, an adequate description of the
acceleration mechanism must include \emph{both} particle and wave
dynamics on an equal footing. In fact the situation is even more difficult,
since the acceleration process turns out to be so efficient that the
pressure of accelerated particles markedly modifies the structure
of the shock (both the overall shock compression and the flow profile).

\section{Accelerated Particles and plasma flow near the shock front}

The transport and acceleration of high energy particles (CRs) near
a CR modified shock is usually described by the diffusion-convection
equation. It is convenient to use a distribution function $f(p)$
normalized to $p^{2}dp$.

\begin{equation}
\frac{\partial f}{\partial t}+U\frac{\partial f}{\partial x}-\frac{\partial }{\partial x}\kappa \frac{\partial f}{\partial x}=\frac{1}{3}\frac{\partial U}{\partial x}p\frac{\partial f}{\partial p}\label{dc1}\end{equation}
 Here $x$ is directed along the shock normal, which for simplicity,
is assumed to be the direction of the ambient magnetic field. The
two quantities that control the acceleration process are the flow
profile $U(x)$ and the particle diffusivity $\kappa (x,p)$. The
first one is coupled to the particle distribution $f$ through the
equations of mass and momentum conservation

\begin{eqnarray}
\frac{\partial }{\partial t}\rho +\frac{\partial }{\partial x}\rho U=0 &  & \label{mas:c}\\
\frac{\partial }{\partial t}\rho U+\frac{\partial }{\partial x}\left(\rho U^{2}+P_{\mathrm{c}}+P_{\mathrm{g}}\right)=0 &  & \label{mom:c}
\end{eqnarray}
 where

\begin{equation}
P_{{c}}(x)=\frac{4\pi }{3}mc^{2}\int _{p_{inj}}^{\infty }\frac{p^{4}dp}{\sqrt{p^{2}+1}}f(p,x)\label{Pc}\end{equation}
 is the pressure of the CR gas and $P_{\mathrm{g}}$ is the thermal
gas pressure. The lower boundary $p_{inj}$ in momentum space separates
CRs from the thermal plasma that is not directly involved in this
formalism but rather enters the equations through the magnitude of
$g$ at $p=p_{inj}$, which specifies the injection rate of thermal
plasma into the acceleration process. The particle momentum $p$ is
normalized to $mc$. The spatial diffusivity $\kappa $, induced by
pitch angle scattering, prevents particle streaming away from the
shock, thus facilitating acceleration by ensuring the particle completes
several shock crossings.

The system (\ref{dc1}-\ref{Pc}) indicate a marked departure from
the test particle theory. Perhaps the most striking result of the
nonlinear treatment is the bifurcation of shock structure (in particular
shock compression ratio) in the parameter space formed by the injection
rate, shock Mach number and particle maximum momentum \cite{mdv00}.

\section{Wave dynamics in the CR shock precursor }

The transformation of magnetic energy to longer scales, while bearing
certain characteristics of the conventional turbulent dynamo problem,
is still rather different from it, in its conventional form. First,
this process should take place in the strongly compressible fluid
near the shock. Second, the Alfven wave turbulence is generated by
accelerated particles via Cerenkov emission, and thus is strongly
coupled to them. Third, the shock precursor itself is unstable to
emission of acoustic waves. The latter phenomenon is known as the
Drury instability and will be discussed later. Acoustic waves, in
turn, interact with particle generated Alfvenic turbulence, stimulating
decay instability (\ie {}{}``inverse cascade'').

\begin{figure}
\begin{center}\includegraphics[  width=0.80\textwidth,
  height=0.40\textheight]{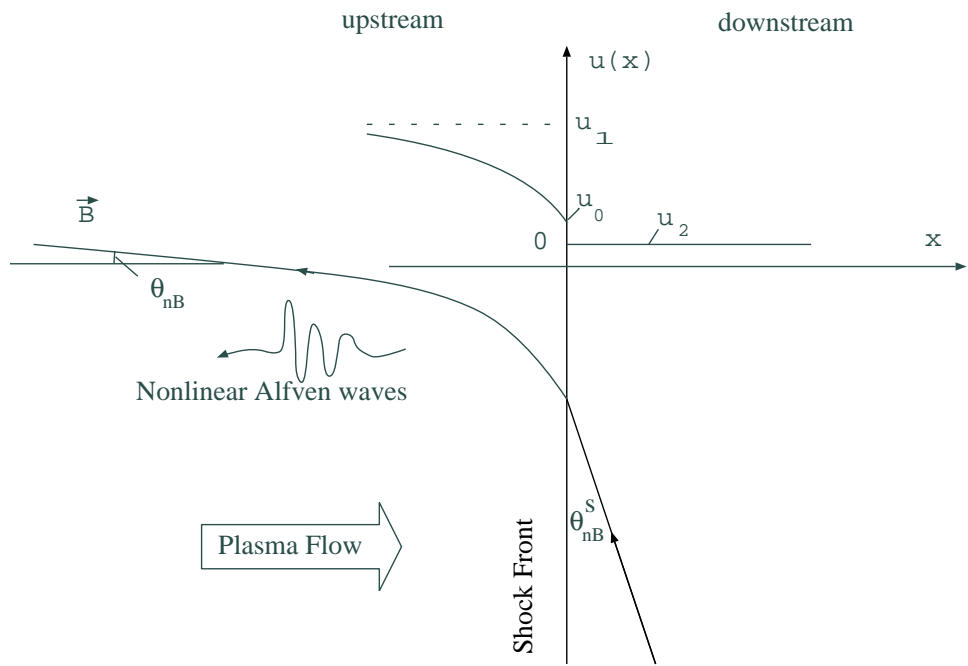}\end{center}

\caption{Schematic representation of nonlinearly accelerating shock. Flow
profile with a gradual deceleration upstream is also shown at the
top.\label{fig:2}}
\end{figure}

The spatial structure of an efficiently accelerating shock, \ie the
shock that transforms a significant part of its energy into accelerated
particles, is very different from that of the ordinary shock, Fig.\ref{fig:2}.
Most of the shock structure consists of a precursor formed by accelerated
CRs diffusing ahead of the shock. If the CR diffusivitity $\kappa (p)$
depends linearly on particle momentum $p$ (as in the Bohm diffusion
case), then, at least well inside the precursor, the velocity profile
$U(x)$ is approximately a \emph{linear} function of $x$, where $x$
points along the shock normal \cite{mdru01}. Ahead of the shock precursor,
the flow velocity tends to its upstream value, $U_{1}$, while on
the downstream side it undergoes a conventional plasma shock transition
to its downstream value $U_{2}$ (all velocities are taken in the
shock frame). This extended CR precursor (of the size $L_{CR}\sim \kappa (p_{max})/U_{1}$)
is the place where we expect turbulence is generated by the CR streaming
instability and where it cascades to longer wavelengths.

\subsection{Alfvenic turbulence}

The growth rate of the ion-cyclotron instability is positive for the
Alfven waves traveling in the CR streaming direction \ie upstream,
and it is negative for oppositely propagating waves. The wave kinetic
equation for both types of waves can be written in the form

\begin{equation}
\frac{\partial N_{k}^{\pm }}{\partial t}+\frac{\partial \omega ^{\pm }}{\partial k}\frac{\partial N_{k}^{\pm }}{\partial x}-\frac{\partial \omega ^{\pm }}{\partial x}\frac{\partial N_{k}^{\pm }}{\partial k}=\gamma _{k}^{\pm }N_{k}^{\pm }+C^{\pm }\left\{ N_{k}^{+},N_{k}^{-}\right\} \label{wke}\end{equation}
 Here $N_{k}^{\pm }$ denotes the number of quanta propagating in
the upstream and downstream directions, respectively. Also, $\omega ^{\pm }$
are their frequencies, $\omega ^{\pm }=kU\pm kV_{A}$, where $V_{A}$
is the Alfven velocity. The linear growth rates $\gamma ^{\pm }$
are nonzero only in the resonant part of the spectrum, $kr_{g}(p_{max})>1$.
In the most general case, the last term on the r.h.s. represents nonlinear
interaction of different types of quanta $N_{k}^{+}$ and $N_{k}^{-}$
and, if compressibility effects are present, also interactions between
the same type. As seen from this equation, the coefficients in the
wave transport part of this equation (l.h.s.) depend on the parameters
of the medium through $U$ and $V_{A}$, which in turn, may be subjected
to perturbations. This usually results in parametric phenomena \cite{sg69}.
We will concentrate on the acoustic type perturbations (which may
be induced by Drury instability), so that we can write for the density
$\rho $ and velocity $U$

\[
\rho =\rho _{0}+\tilde{\rho };\; U=U_{0}+\tilde{U}\]
 The variation of the Alfven velocity $\tilde{V}_{A}=V_{A}-V_{A0}$

\[
\tilde{V}_{A}\simeq -\frac{1}{2}V_{A}\frac{\tilde{\rho }}{\rho _{0}}.\]
 For simplicity, we assume that the plasma $\beta <1$ (which is not
universally true in the shock environment) and neglect the variation
of $U$ compared to that of $V_{A}$ in eq.(\ref{wke}). The above
perturbations of $V_{A}$ in turn induce perturbations of $N_{k}^{\pm }$,
so we can write

\[
N_{k}^{\pm }=\left\langle N_{k}^{\pm }\right\rangle +\tilde{N}_{k}^{\pm }\]
 Our goal is to obtain an evolution equation for the averaged number
of plasmons $\left\langle N_{k}^{\pm }\right\rangle $. Averaging
eq.(\ref{wke:av}) we have

\begin{equation}
\frac{\partial }{\partial t}\left\langle N_{k}^{\pm }\right\rangle +\left(U\pm V_{A}\right)\frac{\partial }{\partial x}\left\langle N_{k}^{\pm }\right\rangle -kU_{x}\frac{\partial }{\partial k}\left\langle N_{k}^{\pm }\right\rangle +\frac{\partial }{\partial k}\left\langle kV_{A}\frac{\tilde{\rho }_{x}}{\rho _{0}}\tilde{N}_{k}^{\pm }\right\rangle =\gamma _{k}^{\pm }\left\langle N_{k}^{\pm }\right\rangle +\left\langle C\left(N_{k}^{\pm }\right)\right\rangle \label{wke:av}\end{equation}
 Here the index $x$ stands for the $x$-derivatives. To calculate
the correlator $\left\langle \frac{\tilde{\rho }_{x}}{\rho _{0}}\tilde{N}_{k}^{\pm }\right\rangle $in
the last equation, we expand the r.h.s. of eq.(\ref{wke}) retaining
only the main linear part in $\tilde{N}$

\begin{equation}
\gamma _{k}^{\pm }N_{k}^{\pm }+C^{\pm }\left\{ N_{k}^{+},N_{k}^{-}\right\} \simeq -\Delta \omega _{k}^{\pm }\tilde{N}_{k}^{\pm }\label{expans}\end{equation}
 The time scale separation between the l.h.s. and r.h.s. of eq.(\ref{wke})
suggests that to lowest order, the linear growth $\gamma ^{+}$ rate
is approximately balanced by the local nonlinear term $C^{+}$. Likewise,
the linear damping of the backward waves $\gamma ^{-}$may be balanced
by their nonlinear growth and conversion of the forward waves $C^{-}$.
Generally, the $\Delta \omega _{k}^{\pm }$ in eq.(\ref{expans})
is a $2\times 2$ matrix operator. If the wave collision term is quadratic
in $N$, then $\Delta \omega _{k}^{\pm }\simeq \gamma _{k}^{\pm }$.
This balance can be established only for the resonant waves ($\gamma ^{\pm }\neq 0$),
whereas our primary focus will be on the extended longwave interval
$k<1/r_{g}(p_{max})$ for which $\gamma \approx 0$. In this domain,
cascading from the generation region $k>1/r_{g}(p_{max})$ takes place
and the refraction (last) term on the l.h.s. of eq.(\ref{wke:av})
plays a dominant role along, with the nonlinear term on the r.h.s.

To calculate the refraction term we write eq.(\ref{wke}), linearized
with respect to $\tilde{N}_{k}^{\pm }$, as:

\begin{equation}
L^{\pm }\tilde{N}_{k}^{\pm }=-kV_{A}\frac{\tilde{\rho }_{x}}{2\rho _{0}}\frac{\partial }{\partial k}\left\langle N_{k}^{\pm }\right\rangle \label{L:oper}\end{equation}
 where

\[
L^{\pm }=\frac{\partial }{\partial t}+\left(U\pm V_{A}\right)\frac{\partial }{\partial x}-kU_{x}\frac{\partial }{\partial k}+\Delta \omega _{k}^{\pm }\]
 Solving eq.(\ref{L:oper}) for $\tilde{N}_{k}^{\pm }$, from eq.(\ref{wke:av})
we thus have the following equation for $\left\langle N_{k}^{\pm }\right\rangle $

\begin{equation}
\frac{\partial }{\partial t}\left\langle N_{k}^{\pm }\right\rangle +U\frac{\partial }{\partial x}\left\langle N_{k}^{\pm }\right\rangle -kU_{x}\frac{\partial }{\partial k}\left\langle N_{k}^{\pm }\right\rangle -\frac{\partial }{\partial k}D\frac{\partial }{\partial k}\left\langle N_{k}^{\pm }\right\rangle =\gamma _{k}^{\pm }\left\langle N_{k}^{\pm }\right\rangle +\left\langle C\left(N_{k}^{\pm }\right)\right\rangle \label{wke:av2}\end{equation}
 Here we introduced a diffusion operator for the Alfven waves in $k$
space due to \emph{random} refraction by the acoustic perturbations
$\tilde{\rho }$ (via the density dependence of $V_{A}$), \ie

\begin{equation}
D_{k}=\frac{1}{4}k^{2}V_{A}^{2}\left\langle \frac{\tilde{\rho }_{x}}{\rho _{0}}L^{-1}\frac{\tilde{\rho }_{x}}{\rho _{0}}\right\rangle \label{diff:op}\end{equation}
 $D_{k}$ is an example of the well-known phenomenon of induced diffusion.
Transforming to Fourier space, we first represent $\tilde{\rho }$
as

\[
\tilde{\rho }=\sum _{q}\rho _{q}e^{iqx-i\Omega _{q}t}\]
 and note that due to the local Galilean invariance of $L$, we can
calculate its Fourier representation in the reference frame moving
with the plasma at the speed $U(x)$ as:

\begin{equation}
L_{k,q}^{\pm }=\pm iqV_{A}+\Delta \omega _{k}^{\pm }-kU_{x}\frac{\partial }{\partial k}\label{L:op:four}\end{equation}
 Then, eq.(\ref{diff:op}) can be re-written as:

\begin{equation}
D_{k}=\frac{1}{2}k^{2}V_{A}^{2}\sum _{q}q^{2}\left|\frac{\rho _{q}}{\rho _{0}}\right|^{2}\Re L_{k,q}^{-1}\label{diff:op2}\end{equation}
 The last (wave refraction) term on the r.h.s. of eq.(\ref{L:op:four})
can be estimated as $U_{1}^{2}/\kappa (p_{max})$, which is the inverse
acceleration time and can be neglected as compared to the frequencies
$qV_{A}$ and $\Delta \omega $. Hence, for $\Re L_{k,q}^{\pm -1}$
we have:

\[
\Re L_{k,q}^{\pm -1}\approx \frac{\Delta \omega _{k}^{\pm }}{q^{2}V_{A}^{2}+\Delta \omega _{k}^{\pm 2}}\]
 For further convenience, we introduce here the number of phonons

\[
N_{q}^{s}=\frac{W_{q}}{\omega _{q}^{s}}\]
 where $W_{q}$ is the energy density of acoustic waves (with $\omega _{q}^{s}=qC_{s}$
as their frequency).

\[
W_{q}=C_{s}^{2}\frac{\rho _{q}^{2}}{\rho _{0}}\]
 For $D_{k}$ in eq.(\ref{wke:av2}) we thus finally have

\[
D_{k}=\frac{k^{2}V_{A}^{2}}{2C_{s}^{2}\rho _{0}}\sum _{q}q^{2}\omega _{q}^{s}\frac{\Delta \omega _{k}^{\pm }}{q^{2}V_{A}^{2}+\Delta \omega _{k}^{\pm 2}}N_{q}^{s}\]
 Note that $D_{k}$ represents the rate at which the wave vector of
the Alfven wave random walks due to stochastic refraction. Of course,
such a random walk necessarily must generate larger scales (smaller
$k$), thus in turn facilitating the confinement (to the shock) of
higher energy particles. Thus, confinement of higher energy particles
is a natural consequence of Alfven wave refraction in acoustic wave
generated density perturbations.

\subsection{Acoustic turbulence}

Unlike the Alfvenic turbulence that originates in the shock precursor
from accelerated particles, there are two separate sources of acoustic
perturbations. One is related to parametric \cite{sg69} processes
undergone by the Alfven waves in the usual form of a decay of an Alfven
wave into another Alfven wave and an acoustic wave. The other source
is the pressure gradient of CRs, which directly drives instability.
The latter leads to emission of sound waves due to the Drury instability.
By analogy with eq.(\ref{wke:av}) we can write the following wave
kinetic equation for the acoustic waves:

\[
\frac{\partial }{\partial t}N_{q}+U\frac{\partial }{\partial x}N_{q}-qU_{x}\frac{\partial }{\partial k}N_{q}=\left(\gamma _{q}^{d}+\gamma _{D}\right)N_{q}+C\left\{ N_{q}\right\} \]
 Here $\gamma _{D}$ is the Drury instability growth rate and $\gamma _{q}^{d}$
is that of the decay instability. We first consider the decay instability
of Alfven waves. Note, however, that the combination of Drury instability
and decay instability can lead to generation of mesoscale fields at
a faster than -- exponential rate, by coupling together the Drury
and decay instability processes.

\subsubsection{Decay Instability}

The mechanism of this instability is the growth of the density (acoustic)
perturbations due to the action of the ponderomotive force from the
Alfven waves. This force can be regarded as a radiative pressure term
appearing in the hydrodynamic equation of motion for the sound waves
(written below in the comoving plasma frame)

\[
\frac{\partial V}{\partial t}=-\frac{1}{\rho _{0}}\frac{\partial }{\partial x}\left(C_{s}^{2}\tilde{\rho }+P_{rad}\right)\]
 Eliminating velocity by making use of continuity equation

\[
\frac{\partial \tilde{\rho }}{\partial t}+\rho _{0}\frac{\partial V}{\partial x}=0,\]
 we obtain

\begin{equation}
\frac{\partial ^{2}\tilde{\rho }}{\partial t^{2}}=\frac{\partial ^{2}}{\partial x^{2}}\left(C_{s}^{2}\tilde{\rho }+P_{rad}\right)\label{sound:eq}\end{equation}
 The Alfven wave pressure can be expressed through their energy

\[
P_{rad}=\sum _{k}\omega _{k}\left(\tilde{N}_{k}^{+}+\tilde{N}_{k}^{-}\right)\]
 Using the relation (\ref{L:oper}) between the density perturbations
and the Alfven waves and separating forward and backward propagating
sound waves $\rho ^{\pm }$, we can obtain from eq.(\ref{sound:eq})
the following dispersion relation for the acoustic branch

\[
\omega ^{2}-q^{2}C_{s}^{2}=q^{2}\sum _{k}\frac{\omega _{k}}{2\rho _{0}}iqkV_{A}L_{k,q}^{\pm -1}\frac{\partial }{\partial k}\left\langle N_{k}^{\pm }\right\rangle \]
 or on writing $\omega =\pm qC_{s}+i\gamma ^{\pm }$, we have the
following growth rate of acoustic perturbations

\[
\gamma ^{\pm }=\frac{q^{2}}{4\rho _{0}}\frac{V_{A}}{C_{s}}\sum _{k}k\omega _{k}L_{k,q}^{\pm -1}\frac{\partial }{\partial k}\left\langle N_{k}^{\pm }\right\rangle \]
 Note that the instability requires an inverted population of Alfven
quanta. As they are generated by high energy resonant particles in
a finite domain of $k$ space, such an inversion clearly can occur.

\subsubsection{Drury Instability}

This instability also leads to efficient generation of acoustic waves
and it is driven by the pressure gradient of the CRs in the shock
precursor. The growth rate has been calculated in Ref.\cite{drury86}
(see also \cite{ber87,zank90,kangjr92}), and can be written in the
form:

\begin{equation}
\gamma _{D}^{\pm }=-\frac{\gamma _{C}P_{C}}{\rho \kappa }\pm \frac{P_{Cx}}{C_{s}\rho }\left(1+\frac{\partial \ln \kappa }{\partial \ln \rho }\right)\label{DrGrRate}\end{equation}
 Here $P_{C}$ and $P_{Cx}$ are the CR pressure and its derivative,
respectively, and $\gamma _{C}$ is their adiabatic index. For an
efficiently accelerating shock $\gamma _{C}\approx 4/3$. Note that
we have omitted a term $-U_{x}$ which is related to a simple compression
of wave number density by the flow and should be generally incorporated
into the r.h.s. of eq.(\ref{sound:eq}) (see \cite{drury86}). This
term is smaller by a factor $C_{s}/U$ than the second (destabilizing)
term. The first term is damping caused by CR diffusion, calculated
earlier by Ptuskin \cite{ptuskin81}.

\section{Mechanisms of transfer of magnetic energy to larger scales}

As it follows from the above considerations, there are a variety of
nonlinear processes that can lead to the transfer of magnetic energy
(generated by accelerated particles in form of the resonant Alfven
waves) to longer scales. First, as it can be seen from eq.(\ref{wke:av2})
(last term on the l.h.s.) scattering of the Alfven waves in $k$ -space
due to acoustic perturbations transfers magnetic fluctuations away
from the resonant excitation region to smaller (but also to larger)
$k$. Second, the nonlinear interaction of Alfven waves and magnetosonic
waves represented by the wave collision term on the r.h.s. can be
responsible for such process. It is also well known that in the presence
of nonzero magnetic helicity there is a strong inverse cascade of
magnetic energy \cite{pouq,moffat}. Finally, even in the frame work
of the weak turbulence, induced scattering of Alfven waves on thermal
protons leads to a systematic decrease in the energy of quanta which,
given the dispersion law, means again transformation to longer waves.

Returning to the wave refraction process on acoustic perturbations
generated by the Drury instability, it is important to emphasize the
following. As is seen from the instability growth rate, eq.(\ref{DrGrRate}),
it is proportional to the gradient of $P_{c}$. As the latter should
be increased as a result of instability, through a better confinement
and faster acceleration, this will reinforce the instability, possibly
triggering {}``explosive'' growth. This can significantly contribute
to mechanisms of regulation of $P_{c}$ discussed earlier in \cite{mdv00}.

\section{Conclusions and Discussion}

We have considered a number of possible mechanisms for generation
of large scale magnetic field in front of strong astrophysical shocks.
All these mechanisms are immediate results of the particle acceleration
process. Such generation is necessary to further accelerate particles
well beyond the {}``knee'' energy at $10^{15}eV$, as is suggested
by observations of the CR background spectrum and the wide consensus
on their SNR shock origin. The fact that accelerated CRs constitute
an ample reservoir for the turbulence which is required to further
accelerate them provides a logical basis for our approach. Indeed,
the scenario proposed here may be viewed as a self-regulating enhanced
acceleration process, which ultimately forces the energetic particle
pressure gradient to its marginal point for Drury instability.

A new description of the instability of Alfven wave spectrum to acoustic
modulations is given. Along with the Drury instability this is shown
to provide an efficient mechanism for transformation of magnetic energy
to longer scales. It should be also emphasized that the theoretical
analysis of CR acceleration is a challenging problem in plasma wave
physics, particle kinetics and shock hydrodynamics. It should, and
indeed must, include a self-consistent description of particle transport,
wave dynamics, and shock structure.

\section*{Acknowledgements}

This work was supported by the U.S. DOE under Grant No. FG03-88ER53275.

\pagebreak

\pagebreak

\end{document}